\documentclass{article}
\usepackage{spconf}
\usepackage{url}
\usepackage{graphicx}
\usepackage{cite}
\usepackage{epsf}
\usepackage{float}
\usepackage{times}
\usepackage{color}
\usepackage{cuted}
\usepackage{stfloats}
\usepackage{tabularx}
\usepackage{graphicx}
\usepackage{arydshln}
\usepackage{amsmath}
\usepackage{amssymb}
\usepackage{wasysym}
\usepackage{multirow}
\usepackage{comment}
\usepackage{mathtools}
\usepackage{subeqnarray}
\usepackage{caption}    
\usepackage{subcaption} 
\usepackage{bm}
\usepackage{hyperref}
\usepackage{booktabs} 
\usepackage{enumitem}
\usepackage{array}       
\usepackage{makecell}
\usepackage{siunitx}
\usepackage{mathtools}
\graphicspath{{Figures/}}
\definecolor{my}{RGB}{0, 0, 255}
\makeatletter

\renewcommand\section{%
  \@startsection{section}{1}{0pt}%
    {-5pt} 
    {2pt}   
    {\normalfont\large\bfseries}%
}

\renewcommand\subsection{%
  \@startsection{subsection}{2}{0pt}%
    {-2pt} 
    {2pt}  
    {\normalfont\normalsize\bfseries}%
}
\makeatother

\begin{document}

\begin{sloppy}

\title{\bf Online Neural Fusion of Distortionless Differential Beamformers for Robust Speech Enhancement}
%
\name{
	\begin{tabular}{c}
		{\it Yuanhang Qian$^{1}$, Kunlong Zhao$^{1}$, Jilu Jin$^{2}$, Xueqin Luo$^{2}$, Gongping Huang$^{1}$,} \\
		{\it Jingdong Chen$^{2}$, and Jacob Benesty$^{3}$}
	\end{tabular}
}
\address{$^{1}$Electronic Information School of Wuhan University, 430072, Wuhan, China \\
$^{2}$CIAIC, Northwestern Polytechnical University, Xi'an, Shaanxi 710072, China \\
$^{3}$INRS-EMT, University of Quebec, Montreal, QC H5A 1K6, Canada \vspace{-15pt}
}

\maketitle

\begin{abstract}
Fixed beamforming is widely used in practice since it does not depend on the estimation of noise statistics and provides relatively stable performance, yet a single beamformer cannot adapt to varying acoustic conditions, which limits its interference suppression capability. To address this, adaptive convex combination (ACC) algorithms have been introduced, where the outputs of multiple fixed beamformers are linearly combined to improve robustness. However, ACC often fails in highly non-stationary scenarios, such as rapidly moving interference, since its adaptive updates cannot reliably track rapid changes. To overcome this limitation, we propose a frame-online neural method for multiple beams fusion, which estimates more efficiently the combination weights. Compared to conventional ACC, the proposed method adapts more effectively to dynamic acoustic environments, achieving stronger interference suppression while maintaining the distortionless constraint.

\end{abstract}



\section{Introduction}
\label{Sect-Intro}
In modern audio applications such as human-computer interaction, speech transcription, and multi-party conferencing, clear and intelligible speech signals are critical to ensure system performance~\cite{elko1996microphone, benesty2023microphone, huang2025advances, yoshioka2012making, benesty2008microphone}. Consequently, microphone arrays combined with beamforming algorithms have become a mainstream technology for enhancing signal quality through spatial filtering~\cite{brandstein2001microphone, elko2008microphone, van1988beamforming}.
	
Beamforming methods are primarily categorized into two types: fixed~\cite{van1988beamforming} and adaptive~\cite{cox1987robust,shahbazpanahi2003robust, griffiths1982alternative}. In fixed beamformers, the differential microphone array (DMA) beamformer is widely used in practical systems due to its core advantages of high directivity, compact structure, and frequency-invariant spatial response~\cite{elko2004differential, benesty2016fundamentals}. However, constrained by its static filtering characteristics, fixed beamformers exhibit limited adaptability in dynamic acoustic environments~\cite{jin2021steering}. When confronted with moving interference sources or changing noise scenarios, the preset filters fail to adequately suppress interference and noise, resulting in reduced speech intelligibility. Adaptive beamforming updates filter coefficients based on environmental parameters or signal statistics to ensure robustness in dynamic acoustic environments~\cite{pan2014performance, liu2025beamforming, frost1972algorithm, habets2010mvdr}. However, obtaining reliable noise statistics in real time is challenging in practice, which limits their effectiveness in highly dynamic or complex scenarios~\cite{besson2005Performance, gu2012Robust}. Jin et al. proposed the ACC method~\cite{jin2024beamforming}, which employs convex combination to adaptively weight the filter coefficients of multiple fixed beamformers~\cite{boyd2004convex}, enhancing adaptability to dynamic acoustic scenarios while ensuring robustness. Nevertheless, the ACC method is still fundamentally based on adaptive updates, which limits its ability to cope with rapidly changing acoustic environments and often results in suboptimal weight estimation.

To overcome the aforementioned technical problems, this paper proposes an online neural method for multiple beamformers fusion~\cite{yan2025neural}. Given a set of fixed beamformers, each designed to satisfy the distortionless constraint while imposing null constraints in different interference directions, the proposed online neural approach learns fusion weights to optimally combine their outputs. 
This enables instantaneous adaptation in highly dynamic acoustic environments, such as those containing single or multiple rapidly moving interferences. Simulation results confirm that the proposed method outperforms both individual beamformers and conventional ACC method in speech enhancement performance. Furthermore, it successfully preserves the target speech while suppressing interference and noise, demonstrating superior robustness under non-stationary conditions.

\section{signal model and Problem Formulation}
\label{Sect-SM-PF}
Consider a uniform linear array (ULA) consisting of $M$ microphones, where the spacing between adjacent microphones is $\delta$ and the speed of sound is $c$. Assuming that a far-field acoustic source signal impinges on the ULA from the desired azimuth angle $\theta_\mathrm{s}$, the phase-delay vector can be expressed as
\begin{align}
\mathbf{d}_{\theta_{\mathrm{s}}} \left(\omega \right)
	=\left[ 1 ~~ e^{-\jmath\omega\tau_0\cos\theta_{\mathrm{s}}} ~~  \cdots ~~ 
	e^{-\jmath(M-1)\omega\tau_0\cos\theta_{\mathrm{s}}}  \right]^T,
\end{align}
where $\jmath$ is the imaginary unit, $\omega=2{\pi}f$ is the angular frequency, $f$ is the temporal frequency, $\tau_0 = \delta/c$ represents the time delay between adjacent microphones for a signal arriving from the endfire direction of the array, and the superscript $^T$ denotes the transpose operator. Therefore, we can obtain the short-time Fourier transform (STFT) representation of the observed signal from the microphone array as~\cite{van2002optimum,benesty2008microphone}
\begin{align}
\label{}
\mathbf{y}\left(w,\ell\right)
&=\left[ \begin{array}{cccc}
	Y_1 \left( \omega,\ell \right) & Y_2 \left( \omega,\ell \right) & \cdots & Y_M \left( \omega,\ell \right)
\end{array} \right]^T \nonumber \\
&=X_{\mathrm{s}} \left(w,\ell\right) \mathbf{d}_{\theta_{\mathrm{s}}} \left(\omega \right) + \mathbf{v}\left(w,\ell\right),
\end{align}
where $X_{\mathrm{s}}\left(w,\ell\right)$ is the desired signal, $\mathbf{v}\left(w,\ell\right)$ is the additive noise and interference components, which is defined similarly to $ \mathbf{y}\left(w,\ell\right)$, and $\ell$ indexes the time frame.

Beamforming refers to the process of estimating the desired signal by applying complex weights to the observed signals form the microphones and summing the results. This operation can be mathematically expressed as
\begin{align}
\hspace{-3pt}
\hat{Z}\left(\omega,\ell\right)
&=\mathbf{h}^H\left(\omega\right)\mathbf{y}\left(\omega,\ell\right) \nonumber\\
&=X_{\mathrm{s}}\left(\omega,\ell\right)\mathbf{h}^{H}\left(\omega\right)
\mathbf{d}_{\theta_{\mathrm{s}}} \left(\omega \right) +\mathbf{h}^{H}\left(\omega\right)\mathbf{v}\left(\omega,\ell\right).
\end{align}
To ensure that the desired signal remains undistorted, beamformers typically impose a distortionless constraint in the desired direction $\theta_{\mathrm{s}}$:
\begin{align}
\mathbf{h}^{H}\left(\omega\right) \mathbf{d}_{\theta_{\mathrm{s}}} \left(\omega \right) =1.
\end{align}

\begin{figure*}[t!]
    \centering
    \includegraphics[width=0.7\textwidth]{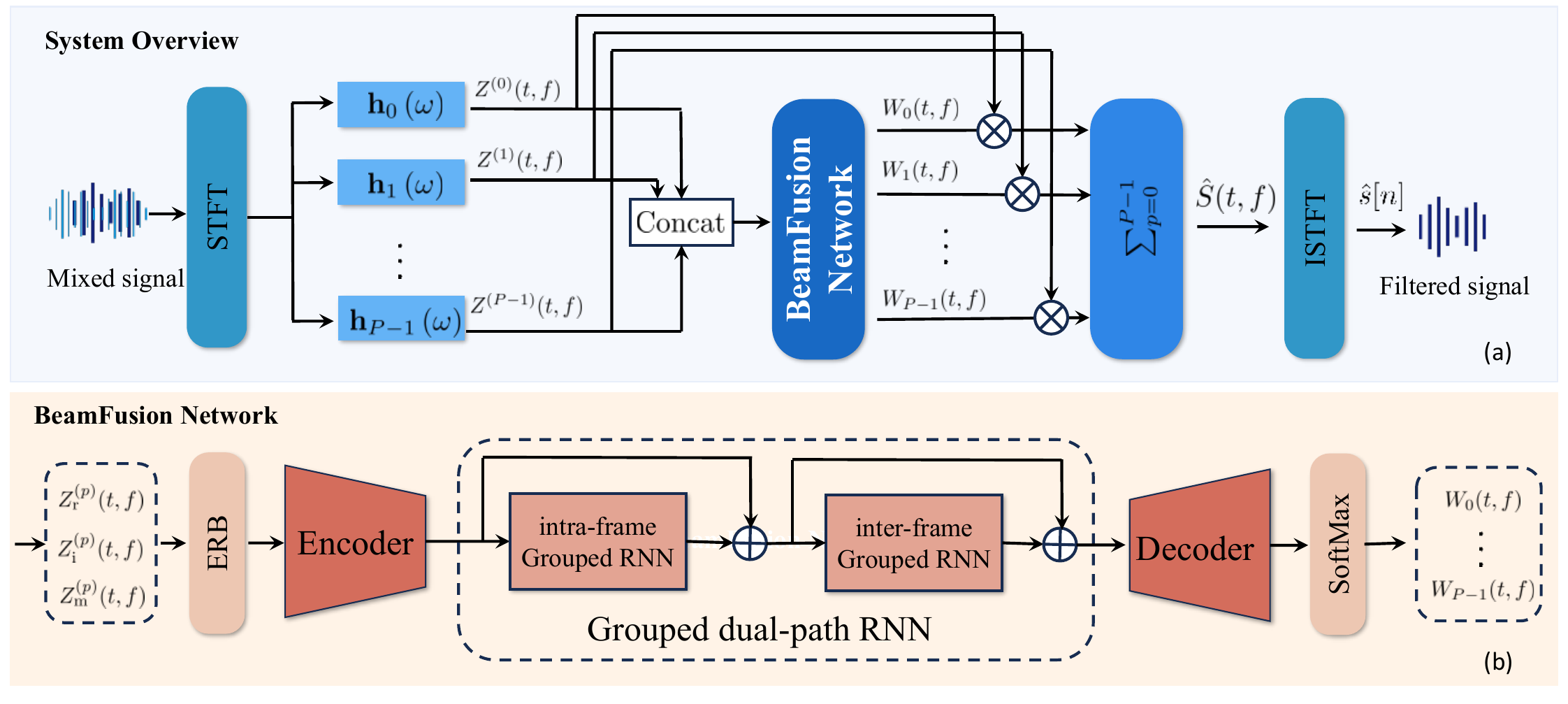}
    \caption{(a)~System overview and (b)~BeamFusion network architecture.}
    \label{fig-network}
    \vspace{-8pt}
\end{figure*}				
\section{Multiple Beamformers Fusion}
\label{Sect-BF}
\subsection{Method Overview}
In practical acoustic environments, multiple types of interference, including fixed or moving sources, as well as background noise, are typically present, making it difficult for a single beamformer to achieve optimal speech enhancement. To address this, ACC algorithms have been developed. Given a set of fixed beamformer coefficients, ACC adaptively combines their weights to produce an optimal estimate of the desired signal.

Define $P$ fixed beamformers $\mathbf{h}_p\left(\omega\right)$, $p=0,1,\dots,P-1$, each steered to the same desired direction $\theta_\mathrm{s}$. The collected filters are organized into a matrix as
\begin{align}
\mathbf{H}\left(\omega\right)=\left[ 
\begin{array}{cccc}
\mathbf{h}_{0}\left(\omega\right)&\mathbf{h}_{1}\left(\omega\right)&\cdots&\mathbf{h}_{P-1}\left(\omega\right)
\end{array}\right].
\end{align}
The objective of this work is to find an optimal filter by linearly combining the existing filters, i.e.,
\begin{equation}
\mathbf{h}_\mathrm{opt}\left(\omega,\ell\right)=\mathbf{H}\left(\omega\right)\bm{\alpha}\left(\omega,\ell\right),
\end{equation}
where 
\begin{align}
\label{alpha}
	\bm{\alpha}\left(\omega,\ell\right) = \left[ \begin{array}{cccc}\alpha_0\left(\omega,\ell\right)~  \alpha_1\left(\omega,\ell\right)
		~ \cdots~ \alpha_{P-1}\left(\omega,\ell\right)
	\end{array}\right]^T
\end{align}
is the combination weight vector, where each element $ 0\leq \alpha_{p}\left(\omega,\ell\right) \leq 1$ for $p = 0,1,\ldots,P-1$.

To ensure that the combined filter $\mathbf{h}_\mathrm{opt}\left(\omega,\ell\right)$ satisfies the distortionless response constraint in the desired direction $\theta_{\mathrm{s}}$, the weight vector $\bm{\alpha}\left(\omega,\ell\right)$ should be restricted:
\begin{align}
\label{eq:distortionless}
\mathbf{h}_\mathrm{opt}^H\left(\omega,\ell\right) \mathbf{d}_{\theta_{\mathrm{s}}} \left(\omega \right) 
&= \sum_{p=0}^{P-1} \alpha_p\left(\omega,\ell\right) \mathbf{h}_p^H\left(\omega\right)  \mathbf{d}_{\theta_{\mathrm{s}}} \left(\omega \right) \nonumber\\
&= \sum_{p=0}^{P-1} \alpha_p\left(\omega,\ell\right) = 1.
\end{align}
The final estimated signal can be expressed as
\begin{align}
\vspace{-6pt}
\hat{{Z}}_{\mathrm{opt}}\left(\omega,\ell\right)
&=\bm{\alpha}^T\left(\omega,\ell\right)\mathbf{H}^H\left(\omega\right)\mathbf{y}\left(\omega,\ell\right)\nonumber\\
&=\sum^{P-1}_{p=0} \alpha_{p}(\omega,\ell) \hat{Z}_{p}(\omega,\ell),
\end{align}
where $\hat{Z}_{p}(\omega,\ell)$ denotes the output of the $p$th fixed beamformer.

In ACC method, the combination weight $\bm{\alpha}\left(\omega,\ell\right)$ is updated using an exponential gradient~\cite{jin2024beamforming,benesty2004exponentiated} approach, which relies on the gradient information of the current frame. While this method can effectively fuses multiple beamformer outputs, its performance is often suboptimal in highly non-stationary scenarios, such as multiple speakers or moving interference sources. In such cases, the gradient estimates may be delayed or inaccurate, preventing null-constrained beamformers from effectively aligning with the interference direction and leading to suboptimal performance. To address this limitation, we propose a neural approach that predicts optimal combination weights.

\subsection{Network Framework}
The overall data flow is illustrated in Fig.~\ref{fig-network}. Given multichannel input signals, we first apply $P$ fixed beamformers to capture acoustic information from different spatial perspectives. The outputs are denoted as $z^{(p)}\left[n\right]$, whose STFT-domain representations are formulated as
$Z^{(p)}\left(t,f\right)$, 
where $p=0,1,\dots,P-1$ refers to the $p$th beamformer, $t$ indexes $T$ frames, and $f$ indexes $F$ frequencies. Then, we extract the real, imaginary, and T-F magnitude spectrum components of the beamformer outputs as $Z_\mathrm{r}^{(p)}\left(t,f\right)$, $Z_\mathrm{i}^{(p)}\left(t,f\right)$, and $Z_\mathrm{m}^{(p)}\left(t,f\right)$, respectively, where the network input $\mathbf{Z}_{\text{in}}\in \mathbb{R}^{3P \times T \times F}$ is constructed by concatenating these three components.

We adopt the network proposed in~\cite{Rong2024GTCRN} as the backbone of our framework. To begin, an equivalent rectangular bandwidth (ERB) filter bank compresses the high-frequency part of $\mathbf{Z}_{\text{in}}$, reducing its frequency dimension from $F$ to $F'$. The compressed features are then encoded into a high-dimensional T-F embedding by an encoder. Subsequently, a grouped dual-path RNN (G-DPRNN) processes the embedding to alternately capture temporal and spectral dependencies. Specifically, the intra-frame RNN models the correlations among frequency bins within a single frame, while the inter-frame RNN characterizes the temporal evolution of each frequency bin across consecutive frames~\cite{Luo2020Dual}. To ensure causality, this module employs unidirectional GRUs. The following decoder generates the weight $W_p\left(t,f\right)$, which is equivalent to the coefficient $\alpha_p\left(\omega,\ell\right)$ defined in~(\ref{alpha}). It is worth mentioning that a softmax operation is applied to the weights before output, enforcing the distortionless constraint $\sum_{p=0}^{P-1} W_p\left(t,f\right) = 1$ in~(\ref{eq:distortionless}).
The estimated signal is obtained by a weighted fusion of the beamformer outputs:
\begin{equation}
\vspace{-3pt}
\hat{S}(t,f) = \sum_{p=0}^{P-1} W_p(t,f)\, Z^{(p)}(t,f).
\vspace{-3pt}
\end{equation}
Eventually, the time-domain signal $\hat{s}[n]$ is obtained by applying the inverse STFT to $\hat{S}(t,f)$.

In the design of the loss function, the primary constraint is that the output of the weighted enhanced signal should approximate the original direct sound component as closely as possible, ensuring the fidelity of the target speech. Following this guideline, the loss function is formulated as
\begin{equation}
\mathcal{L} = \mathbb{E}
\left[ \left\| \mathbf{X}_\mathrm{fi} - \mathbf{X}_{\mathrm{ref}} \right\|_{2}^{2} \right],
\end{equation}
where $\mathbf{X}_{\mathrm{fi}}$ represents the filtered desired signal, $\mathbf{X}_{\mathrm{ref}}$ denotes the reference desired signal,  $\mathbb{E}[\cdot]$ is the mathematical expectation, and $\left\|\cdot\right\|_{2}$ is the Euclidean norm. By minimizing this loss, the network learns adaptive fusion weights that preserve the integrity of the target speech. For brevity, we refer to the proposed method as BeamFusion.

\section{Simulations}
\label{Sect-Exp}
\subsection{Data Preparation}
An 8-element ULA with an inter-element spacing of $\delta = 1~\mathrm{cm}$ is employed. The array is placed in a rectangular room of size ${8~\mathrm{m} \times 6~\mathrm{m} \times 3~\mathrm{m}}$, oriented parallel to the $x$-axis, with its center located at coordinates $(4, 2, 1)~\mathrm{m}$. The target source is fixed at an azimuth of $\theta_{\mathrm{s}}=0^\circ$, positioned at $(6, 2, 1)~\mathrm{m}$. The interfering source is placed at a distance of $2~\mathrm{m}$ from the array center, initially located at an azimuth of $90^\circ$, and rotated counterclockwise around the array center by $10^\circ$ every second until reaching $180^\circ$.

Speech signals are selected from the LibriSpeech corpus~\cite{panayotov2015librispeech} and segmented into $10~\mathrm{s}$ clips, with silence removal applied to ensure that no segment contained a continuous silent period longer than $0.1~\mathrm{s}$. Room impulse responses (RIRs) are generated using the image source method~\cite{allen1979image,pan2023anchor}, with reverberation time $T_{60} \in \left[200, 800\right]~\mathrm{ms}$. For each simulated scenario, target and interfering sources are randomly selected from the prepared clips. Their signals are convolved with the corresponding RIRs, combined, and subsequently contaminated with Gaussian white noise at a randomly chosen signal-to-noise ratio (SNR) between $\SI{20}{dB}$ and $\SI{40}{dB}$. Following this procedure, a total of $10{,}000$ audio samples are generated for training and $600$ samples for testing.

\subsection{Beamformers Configuration}
In this study, seven beamformers are investigated for comparison. 
\begin{itemize}[itemsep=0pt, topsep=0pt, parsep=0pt, partopsep=0pt]
\item MWNG beamformer~\cite{huang2022fundamental}.
\item Four first-order differential beamformers~\cite{benesty2012study}, denoted as DMA-\textrm{I}, DMA-\textrm{II}, DMA-\textrm{III}, and DMA-\textrm{IV}. All of them are steered towards the desired direction at $\theta_{\mathrm{s}} = 0^\circ$, with nulls placed at $90^\circ$, $120^\circ$, $150^\circ$, and $180^\circ$, respectively. 
\item The ACC method~\cite{jin2024beamforming}, which adaptively fuses the outputs of the above beamformers.
\item The proposed BeamFusion method, which employs a neural network to estimate the fusion weights of the above beamformers.
\end{itemize}

\subsection{Model Configuration}
In the feature processing stage, the input signals are first transformed using the STFT, where the fast Fourier transform (FFT) size is set to $512$, the length of window $512$, and the overlap $128$. We set the encoder output, decoder input, and G-DPRNN hidden dimension to $32$, while all other parameters remain the same as in~\cite{Rong2024GTCRN}. The total number of parameters of the model is $85$~k, with a computational complexity of $142$ M Macs.

The model is trained using the Adam optimizer~\cite{kingma2014adam} with an initial learning rate of $1.0 \times 10^{-3}$. Whenever the validation loss did not decrease for $5$ consecutive epochs, the learning rate is reduced by half, with a minimum allowed learning rate of $1.0 \times 10^{-4}$. The batch size is set to $30$, and the training process is conducted for a total of $50$ epochs.

During inference, the network performs frame-level online processing, where fusion weights are predicted from the current and past frames and the resulting outputs are overlap-added with nearby frames to reconstruct the final signals.

\begin{table}[t!]
    \centering
    \caption{Performance comparison under different reverberation conditions with moving interference.}
    \label{tab:reverb}
    \renewcommand{\arraystretch}{1.1}
    \setlength{\tabcolsep}{2.5pt}
    \scalebox{0.7}{%
    \begin{tabular}{lcccccc}
        \toprule
        & \multicolumn{3}{c}{$T_{60} = 300~\mathrm{ms}$} 
        & \multicolumn{3}{c}{$T_{60} = 700~\mathrm{ms}$} \\
        \cmidrule(lr){2-4} \cmidrule(lr){5-7}
Method & \makecell{$\triangle$SNR~(dB)} & STOI & \makecell{$\triangle$SI-SDR~(dB)} 
       & \makecell{$\triangle$SNR~(dB)} & STOI & \makecell{$\triangle$SI-SDR~(dB)} \\
        \midrule
        Noisy        & $-$    & $0.58$ & $-$    & $-$    & $0.44$ & $-$ \\
        MWNG         & $0.28$ & $0.62$ & $0.22$ & $0.28$ & $0.46$ & $0.19$ \\
        DMA-I        & $5.31$ & $0.69$ & $5.50$ & $6.41$ & $0.56$ & $6.83$ \\
        DMA-II       & $6.02$ & $0.71$ & $6.16$ & $6.04$ & $0.55$ & $6.34$ \\
        DMA-III      & $4.58$ & $0.69$ & $4.68$ & $4.28$ & $0.51$ & $4.48$ \\
        DMA-IV       & $4.95$ & $0.69$ & $5.06$ & $4.68$ & $0.52$ & $4.91$ \\
        ACC          & $6.56$ & $0.73$ & $6.61$ & $7.08$ & $0.58$ & $7.30$ \\
        BeamFusion   & $\bm{10.24}$ & $\bm{0.74}$ & $\bm{7.13}$ 
                     & $\bm{12.07}$ & $\bm{0.58}$ & $\bm{7.58}$ \\
        \hline
        \bottomrule
    \end{tabular}
    }
\end{table}

\begin{figure}[t!]
	\vspace{8pt}
	\centerline{\includegraphics[width=70mm]{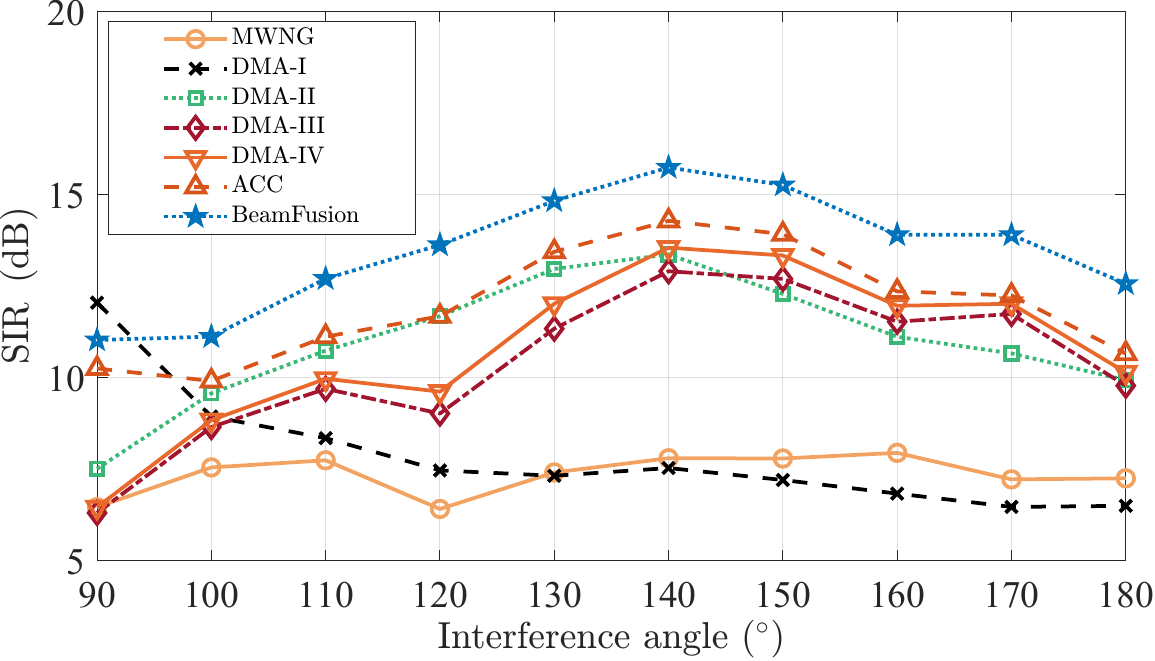}}
	\caption {SIR performance of MWNG, DMAs, ACC, and BeamFusion methods across different interference angles from $90^\circ$ to $180^\circ$.}
	\label{Fig2_exp2}
	\vspace{-12pt}
\end{figure}

\section{Results and Discussion}
In the first experiment, we evaluate the performance of the proposed BeamFusion method in a moving interference scenario under different reverberation conditions. The compared methods include the unprocessed noisy signal, MWNG, DMA-\textrm{I}, DMA-\textrm{II}, DMA-\textrm{III}, DMA-\textrm{IV}, and ACC method. To assess the robustness of the different approaches, different reverberation times are considered, corresponding to $T_{60}=300~\mathrm{ms}$ and $700~\mathrm{ms}$. The evaluation metrics include the improvements in signal-to-noise ratio ($\triangle$SNR) and scale-invariant signal-to-distortion ratio ($\triangle$SI-SDR), as well as the short-time objective intelligibility~\cite{STOI2010} (STOI). The results are summarized in Table~\ref{tab:reverb}. The ACC method improves $\triangle$SNR and speech intelligibility by adaptively combining multiple beamformers, outperforming any single DMA. The proposed BeamFusion approach surpasses both ACC and individual DMAs across all metrics, with particularly significant gains in $\triangle$SNR. Its performance remains consistently superior under different reverberation conditions, demonstrating the method’s effectiveness in enhancing the target signal while suppressing interference.

In addition, we evaluate the distribution of the signal-to-interference ratio (SIR) metric as a function of the interference angle for signals processed by BeamFusion and other methods. A total of $200$ simulations were conducted with the reverberation time set to $T_{60}=240~\mathrm{ms}$, and the evaluation metrics were averaged. Since the network structure of BeamFusion is nonlinear, it is difficult to directly extract the filtered interference signal. Therefore, we adopt the BSS evaluation method proposed in~\cite{vincent2006performance} to assess the performance. This method applies an optimal linear filter to align the estimate with the reference, then decomposes it into target, distortion, and interference components. Based on this decomposition, the SIR metric is calculated to quantify the level of interference suppression. Figure~\ref{Fig2_exp2} illustrates the SIR curves of the evaluated methods across different interference angles. As shown in the figure, BeamFusion consistently achieves higher SIR values than the other methods for most interference angles, demonstrating its superior capability in suppressing interfering sources.

\begin{table}[t]
	\centering
	\caption{Performance comparison in multi-interference environment. Conditions: $T_{60} = 300~\mathrm{ms}$.}
	\label{tab:reverb300}
	\renewcommand{\arraystretch}{1.1}
	\setlength{\tabcolsep}{15pt}
    \scalebox{0.8}{%
	\begin{tabular}{lccc}
		\toprule
		Method & $\triangle$SNR~(dB) & STOI & $\triangle$SI-SDR~(dB) \\
		\midrule
		Noisy        & $-$    & $0.49$ & $-$    \\
		MWNG         & $0.45$ & $0.55$ & $0.46$ \\
		DMA-I        & $5.30$ & $0.60$ & $5.40$ \\
		DMA-II       & $7.01$ & $0.66$ & $7.10$ \\
		DMA-III      & $5.42$ & $0.62$ & $5.50$ \\
		DMA-IV       & $5.86$ & $0.63$ & $5.94$ \\
		ACC          & $7.48$ & $0.67$ & $7.52$ \\
		BeamFusion   & $\bm{12.16}$ & $\bm{0.68}$ & $\bm{7.99}$ \\
		\hline
        \bottomrule
	\end{tabular}
    }
    \vspace{-5pt}
\end{table}

Furthermore, we investigated the noise suppression performance of BeamFusion under multi-interference conditions. The experimental setup was identical to the previous configuration, except for the addition of a fixed interference source randomly located between $90^\circ$ and $270^\circ$ in azimuth, with $T_{60}=300~\mathrm{ms}$. Performance metrics were obtained by averaging results across 200 signals. As shown in Table 2, the BeamFusion approach outperforms any individual beamformers as well as the ACC method, demonstrating superior generalization and robustness in unseen acoustic conditions.

\section{Conclusions}
\label{Sect-Conclu}
In this work, we proposed an online neural multiple beamformers fusion framework that integrates the outputs of multiple fixed beamformers through learned weights. Unlike conventional ACC methods that rely on adaptive updates and often provide suboptimal estimates in rapidly changing environments, the proposed framework is capable of realizing the optimal fusion strategy while maintaining the distortionless constraint. Experimental results demonstrate that the proposed framework consistently outperforms both fixed and ACC-based beamformers, offering stronger interference suppression and enhanced speech intelligibility, thereby highlighting its potential for real-time, high-fidelity speech enhancement.

\footnotesize

\bibliographystyle{IEEEtran}
\bibliography{Bib_MABFSE}

\end{sloppy}
\end{document}